\documentstyle[12pt]{article}
\normalbaselines
\begin{document}
quant-ph/9704017
\begin{center}
\vspace*{1.0cm}

{\large{\bf Entropy and optimal decompositions of states }}

{\large{\bf relative to a maximal commutative subalgebra }}
\footnote{Dedicated to Walter Thirring at his 70th birthday}

\end{center}
\vskip 1.5cm

Armin Uhlmann

\vskip 0.5 cm

Institut f\"ur Theoretische Physik,       \hfill \\
Universit\"at Leipzig,                    \hfill \\
Augustusplatz 10,                         \hfill \\
D-04109, Leipzig, Germany.                \hfill \\

\vskip 0.5 cm

\begin{abstract}
To calculate the entropy of a subalgebra or of a channel with respect
to a state, one has to solve an intriguing optimalization problem.
The latter is also the key part in the entanglement of formation concept,
in which case the subalgebra is a subfactor. \hfill \\
I consider some general properties, valid for these
definitions in finite dimensions,
and apply them to a maximal commutative subalgebra
of a full matrix algebra. The main method
is an interplay between convexity and symmetry.
A collection of helpful tools from convex analysis
is collected in an appendix.
\end{abstract}

\vspace{1 cm}

{\bf INTRODUCTION}

This paper considers the {\it entropy of a subalgebra or of a
completely positive map
with respect to a state}, an entropy-like quantity introduced by
A.~Connes, H.~Narnhofer, and W.~Thirring. I remain,
however, within a rather narrow setting:
A pair of algebras, $^*$-isomorphic to the algebra of all
$d \times d$-matrices, and to its subalgebra of diagonal matrices.
I depart from this restriction within this introduction and in discussing
some tools from convex analysis (lemmata 1, 2, 3, and appendix).

While the {\it von Neumann entropy} is of undoubted relevance for type
$I$ algebras (with discrete center), the relative entropy can be
meaningfully defined even on the state space of an arbitrary
$^*$-algebra. There are, depending on the category of algebras and
states, several quite different ways to do so, \cite{OP93}.

In \cite{NT85} Narnhofer and Thirring proposed a von Neumann entropy
definition by the aid of relative entropy. At the end of their paper
they mentioned a quantity now denoted by
$H_{\omega}({\cal A})$ or $H_{\omega}({\cal B}|{\cal A})$ where
${\cal A}$ is a unital subalgebra of ${\cal B}$, and $\omega$ a state
of ${\cal B}$. Abbreviating the restriction of $\omega$ onto
${\cal A}$ by $\tilde \omega$,
$$
 \omega \mapsto \tilde \omega := \omega_{| {\cal A}},
$$
their definition reads
$$
H_{\omega}({\cal B}|{\cal A}) = H_{\omega}({\cal A}) :=
\sup \, \sum \, p_j S(\tilde \omega_j, \tilde \omega),
\qquad {\cal A} \subset {\cal B} \eqno(*)
$$
In this expression $S(. , .)$ is the relative entropy for
the states of ${\cal A}$ and the supremum has to run through
all convex decompositions
$$
 \omega = \sum p_j \omega_j
$$
of the state $\omega$ on ${\cal B}$. ($*$) was later called
``entropy of a subalgebra with respect to a state''.

It depends concavely on the state, is always non-negative,
and it inherits from relative entropy its monotonicity:
$$
{\cal A}_1 \subset {\cal A}_2 \subset {\cal B} \, \longrightarrow
\, H_{\omega}({\cal A}_1) \leq H_{\omega}({\cal A}_2) \leq S(\omega)
$$
According to \cite{NT85} $S(\omega) = H_{\omega}({\cal B} | {\cal B})$
is the von Neumann entropy of $\omega$.

($*$) amounts to calculate a number. Seeing the ease and elegance
of the definition one might perhaps not believe what
a formidable task this is. The calculational
difficulties are mainly encoded in $R$, a functional
defined by
$$
H_{\omega}({\cal B}|{\cal A}) + R({\cal B}|{\cal A}, \omega)
= H_{\omega}({\cal A}|{\cal A}) \equiv S(\tilde \omega)
$$
where $\omega$ is a state of ${\cal B}$.

All terms are non-negative. If $S(\tilde \omega) < \infty$
they are finite. $R$ is the convex hull of the
function $\omega \to S(\tilde \omega)$,
as explained below. In the finite dimensional case
one may write
$$
R({\cal B}|{\cal A}, \omega) = \inf \sum p_j S(\tilde \varrho_j),
$$
the infimum being taken over all extremal convex combinations
$$
 \omega = \sum p_j \varrho_j, \quad \varrho_k \hbox{ pure}
$$
Therefore, $R$ is already determined by its values at pure states.

In the present paper ${\cal A}$ is a maximal commutative subalgebra.
The case of a general subalgebra has been considered by
Benatti, Narnhofer, and Uhlmann, \cite{BNU96}. But their
main examples concern maximal commutative subalgebras. The same
is with the paper of Benatti, \cite{Ben96}, who shows
a relation of ($*$) to {\it accessible entropy}.
An example with another subalgebra is in \cite{Uh97a}.

Results and completely independent definitions for the subfactor
problem are due to
Bennett, DiVincenzo, Smolin, and Wootters, \cite{BDSW96}, who
aimed at the $R$-function on the states of a direct product
of two finite dimensional factors. They
defined the {\it entanglement of formation} by
$$
{\cal A} \hbox{ a subfactor } \longrightarrow
E(\omega) \equiv R(\omega) = \hbox{ entanglement of } \omega
\hbox{ with respect to } {\cal A}
$$
in order to measure entanglement as a resource for quantum information
transfer. They obtained estimates and solved interesting examples.
Because the reductions of a pure state to a factor and to its commutant
have the same entropy, there is a nice symmetry
$$
R({\cal B}_1 \otimes {\cal B}_2 | {\cal B}_1, \omega) =
R({\cal B}_1 \otimes {\cal B}_2 | {\cal B}_2, \omega)
$$
Hill and Wootters, \cite{HW97}, solved the problem for rank two
states on the direct product of two 2-dimensional matrix algebras.
\footnote{Meanwhile, B.W.Wootters has solved the 2-qubit case
completely, quant-ph/9709029 v2.}

The definition ($*$) can be extended to a completely positive unital
map, $\alpha$, from one algebra to another one,
$$
\alpha \, : \quad {\cal A} \mapsto {\cal B}
$$
Its transpose, a stochastic mapping,
$$
\omega \mapsto \omega \circ \alpha, \quad (\omega \circ \alpha)(A)
= \omega(\alpha(A)), \, \, A \in {\cal A}
$$
maps states of ${\cal B}$ to those of ${\cal A}$.

To get the definition one has only to set
$\tilde \omega := \omega \circ \alpha$ within ($*$) to obtain
the desired quantity $H_{\omega}(\alpha)$. This is an invention
of Connes, Narnhofer, and Thirring in \cite{CNT87}.

Ohio and Petz called $\alpha$ a {\it channel map} acting
from the {\it output algebra}
${\cal A}$ to the {\it input algebra} ${\cal B}$, so that its
transpose, a stochastic mapping, acts from the states of the
input algebra into the state space of the output one.
In their monograph \cite{OP93}, in which they
consider the problem within the C$^*$- and the W$^*$-category,
$H_{\omega}(\alpha)$ is called {\it entropy of the channel
$\alpha$ with respect to the state} $\omega$. Now the
monotonicity property reads
$$
H_{\omega}(\alpha_1 \circ \alpha_2) \leq  H_{\omega}(\alpha_1)
\leq S(\omega)
$$
whenever $\alpha_1 \circ \alpha_2$ is well defined, unital, and
completely positive. As known, the latter requirement can be weakened
to Schwarz positivity
$$
\alpha(A^* A) \geq | \alpha(A) |^2 \quad \forall \, A \in {\cal A}
$$
$$     $$

{\bf GENERAL PROPERTIES}

Let ${\cal H}$ be a Hilbert space of finite dimension
$\dim {\cal H} = d$, and ${\cal C}$ a maximal commuting subalgebra
of ${\cal B} :={\cal B}({\cal H})$. Let us denote by
$P_j = |j \rangle \langle j|$, $ j = 1, 2, \dots, d$, the minimal
projection operators of ${\cal C}$. They support the  distinguished
pure states $\varrho^C_j$, i.~e.
\begin{equation} \label{th01}
P_j B \, P_j = \varrho^C_j(B) \, P_j, \quad \forall \, \, B \in {\cal B}
\end{equation}

The density matrix of a state of ${\cal B}$ is contained in ${\cal C}$
iff the latter is a convex combination of the pure states $\varrho_k^C$.
The restriction $\tilde \omega$ of a state $\omega$ onto
${\cal C}$ can hence be described by the {\it reduction map}
\begin{equation} \label{th02}
\omega \rightarrow \tilde \omega := \sum \omega(P_j) \, \varrho^C_j
\end{equation}

Now we consider entropies. All what is needed is nicely reviewed
in \cite{Weh78}. The entropy of the restriction $\tilde \omega$ of
$\omega$ onto ${\cal C}$ reads

\begin{equation} \label{th03}
\tilde S(\omega) := S(\tilde \omega) = - \sum \omega(P_j)
\, \ln( \omega(P_j) )
\end{equation}

It is now possible to write down the
{\it entropy of ${\cal C}$ with respect of a state} $\omega$
{\it of} ${\cal B}$ as defined by Narnhofer and Thirring \cite{NT85},
Connes \cite{Co85}, and \cite{CNT87}.
In the case at hand the general definition is equivalent to

\begin{equation} \label{HN2}
H_{\omega}({\cal C}) : =
\tilde S(\omega) - R(\omega), \quad
R(\omega) = \inf \sum p_j \tilde S(\omega_j)
\end{equation}

where the infimum runs through all convex decompositions

\begin{equation} \label{conv01}
 \omega = \sum p_j \omega_j
\end{equation}

in the state space $\Omega$ of ${\cal B}$. Rockafellar \cite{Roc70}
calls the construction used in defining $R$ {\it the convex hull of}
$\tilde S$. The convex hull of any function on any convex set is
always convex. Thus $H_{\omega}$ is the sum of two concave functions,
$\tilde S$ and $-R$, and hence concave.

For the following it is essential that $R$ is the convex hull of
a {\it concave} function, and that $\Omega$ as well as
$\Omega^{\rm ex}$, the set of its extremal points, are compact.
Being the state space of ${\cal B}$, a state is extremal iff
it is pure. A state $\varrho$ is pure iff there is a projection
operator $P \in {\cal B}$, the support of $\varrho$, such that
$PBP = \varrho(B) \, P$ for all $B$ in that algebra.
Notice: {\it lemmata 1, 2, and 3 below are valid for every
unital subalgebra}, the restriction to a maximal commutative
subalgebra is not essential for their validity.

The first conclusion is due to the concavity of $\tilde S$:
It is possible to restrict (\ref{conv01})
to {\it extremal convex decompositions},

\begin{equation} \label{conv02}
 R(\omega) = \inf \sum p_j \tilde S(\varrho_j), \quad \varrho_k \in
\Omega^{\rm ex}, \quad \sum p_j \varrho_j = \omega
\end{equation}

Let us call {\it optimal} every extremal convex
decomposition of $\omega$ with which the infimum
(\ref{conv02}) is attained, and for which $p_k > 0$ for all its
coefficients \cite{BNU96}. Thus optimality is expressed by

\begin{equation} \label{opt01}
 R(\omega) = \sum p_j \tilde S(\varrho_j), \quad \varrho_k \in
\Omega^{\rm ex}, \quad \forall \, p_k > 0
\end{equation}

The graph of $R$ is a closed subset of the boundary of a compact
convex set, \cite{Uh97a}, see appendix. It implies, by standard
arguments, that there are optimal decompositions for every state.
Then, according to Carath\'eodory, there exist simplicial ones.
This is the content of

{\bf Lemma 1}

Every $\omega$ admits an optimal decomposition with at least
rank$(\omega)$ and at most rank$(\omega)^2$ different pure
states. $\Box$

I need some further, almost obvious conclusions from the
definition of $R$. For the time being a convex subset
$\Omega_0$ of $\Omega$
will be called an $R${\it -set} if every $\omega \in \Omega_0$
admits an optimal decomposition into pure states of $\Omega_0$.
It is clear that

a) every $R$-set $\Omega_0$ of the
state space is the convex hull of its pure states.
(Therefore, the pure states contained in $\Omega_0$ are just
the extremal elements of $\Omega_0$.)

b) that $R$, restricted to $\Omega_0$, can be computed by
optimal decompositions (\ref{opt01}) into pure states
which are all contained in $\Omega_0$,

c) and that every face of $\Omega$ is an $R$-set.

{\bf Lemma 2}

Let $\Omega_0$ be an $R$-set and $\Omega_0^{\rm ex}$
the set of its pure states. Let $F$ be a convex function on
$\Omega_0$ which is not greater than $R$ on $\Omega_0^{\rm ex}$.
Then $F \leq R$ on $\Omega_0$. $\Box$

With other words, on any $R$-set $\Omega_0$
of the state space, $R$ is the largest convex function
which attains at every of its pure states, $\varrho$,
the value $\tilde S(\varrho)$. Indeed, for an optimal decomposition,
based on $\Omega_0$, convexity of $F$ implies
$$
R(\omega) = \sum p_j \tilde S(\varrho_j) \geq
\sum p_j F(\varrho_j) \geq F(\omega)
$$

My next task is to apply this simple lemma to affine functions
in order to obtain a slight modification of theorem 1 of \cite{BNU96}:
$\Omega$ can be covered by convex sets on which $R$ is
affine. This fact I like to call the {\it roof property}
of $R$.
The covering consists of ``facets'' with pure states as
corners. The covering is not a disjunct one. The intersection
of two ``facets'' is either empty or belongs itself to the
covering.

To obtain the covering, we use the well known possibility to
represent a convex function on a compact convex domain by
an upper bound of affine functions, together with the existence
of optimal decompositions.

An affine function, $l$, $\omega \to l(\omega)$ is said to {\it support}
$R$ iff $l \leq R$ on $\Omega$, and $l$ equals $R$ at least at one state.
By virtue of lemma 2 one needs to check the inequality $l \leq R$
for pure states only.

$\Omega$ is compact and $R$ convex. Hence there exists for
every $\omega' \in \Omega$ at least one affine function $l$
supporting $R$ at $\omega'$, i.~e.~with $l(\omega') = R(\omega')$.
Thus
$$
 R(\omega) = \sup_l l(\omega), \quad l \, \hbox{ supports } \, R
$$
Now, if $l$ is supporting $R$, let us consider the set

\begin{equation} \label{conv03}
\Omega(l) := \{ \,  \omega' \in \Omega \, | \, \, R(\omega') =
l(\omega') \, \, \}
\end{equation}

on which $l$ coincides with $R$, and let us assume, $\omega$
belongs to that set.  Choosing an
optimal decomposition (\ref{opt01}) one obtains
$$
\sum p_j R(\varrho_j) = R(\omega) = l_{\omega}(\omega)
= \sum p_j l_{\omega}(\varrho_j)
$$
But $l_{\omega}(\varrho_j) \leq R(\varrho_j)$ as $l$ is
$R$-supporting.
The positivity of the coefficients $p_j$ enforces
equal values of $R$ and $l$ for all involved pure states.
This is not the end: $l$ is affine and equal to $R$ on some
extremal elements $\varrho_j$. Therefore, by convexity, $R \leq l$
on the convex hull of the pure states $\varrho_j$. But $l \leq R$
by assumption. Hence $l$
is equal to $R$ on the convex hull of all the pure states
$\varrho$ which can appear in any optimal decomposition of
$\omega$. This is already the essence of

{\bf Lemma 3}

Let $\Omega(l)$ be defined by (\ref{conv03}) with an affine function
$l$ supporting $R$. Then $\Omega(l)$ is a compact, convex
$R$-set on which $R$ is affine.

The family of all $\Omega(l)$, where $l$ is $R$-supporting,
is a covering of $\Omega$.
$\Box$

{\it Proof:}

Up to the compactness assertion the proof is already done by the
chain of arguments above, which can be repeated with {\it every}
$\omega \in \Omega(l)$. In particular, $\Omega(l)$ is an $R$-set.
Now $R$ equals $\tilde S$ on the compact
set $\Omega^{\rm ex}$. Hence both, $R$ and $l_{\omega}$ are
continuous on this compact set. Hence,  the subset of
$\Omega^{\rm ex}$, on which both functions take equal values,
is compact. This compact set of extremal points generates a
compact convex set (Carath\'eodory) which must be $\Omega(l)$
as it is an $R$-set. $\Box$

{\it Remark:}

${\cal B}$ being finite-dimensional, the Hermitian linear
functionals span a real linear space. An affine function on it
is of the form $l(\nu) = \nu(A_l) + a$ with a real constant $a_l$
and with an Hermitian operator $A_l$. For every state,
$\omega$, the constant $a$ can be represented by $a$-times
the evaluation of $\omega$ at the identity.
Hence, for every affine $l$ supporting
$R$, there is exactly one Hermitian operator $A$ such that the
expectation value $\omega(A)$ equals $l(\omega)$,

\begin{equation} \label{dualform}
l(\omega) = \omega(A_l), \quad A_l = A_l^* \in {\cal B}
\end{equation}

With an Hermitian operator $A$, satisfying
$\tilde S(\varrho) \geq \varrho(A)$  for all pure states $\varrho$,
the expectation functional $A \to \omega(A)$ satisfies
$R(\omega) \geq \omega(A)$ on the whole state space by virtue
of lemma 2. If, in addition, equality takes place on at least
one pure state, then the expectation value of $A$ is supporting
$R$. I shall consider this aspect elsewhere. $\Box$

There may be many linear functionals $l$ supporting $R$ at a
given state $\omega$. For every pure state $\varrho$,
appearing in any extremal optimal decomposition of $\omega$,
we get $l(\varrho) = R(\varrho)$ and $\varrho \in \Omega_l$.
Hence

{\bf Corollary}

Let $\omega$ be a state. The intersection

\begin{equation} \label{facetdef}
\Omega_{\omega} := \bigcap \, \Omega(l), \quad l(\omega) = R(\omega)
\end{equation}

enjoys the following properties: It is convex, compact, and it contains
every pure state which can appear in an optimal decomposition of
$\omega$. $R$, restricted to $\Omega_{\omega}$, is affine. $\Box$

$\Omega_{\omega}$ is a simplex iff $\omega$ allows for one and
only one extremal optimal decomposition (up to the order of
its summands).

{\it Remark:} In \cite{Uh97a} I have called $\Phi_{\omega}$
the convex set generated by those pure states in
$\Omega_{\omega}$ which can appear in an extremal optimal
decomposition of $\omega$. Clearly, $\Phi_{\omega}$ is contained
in $\Omega_{\omega}$. Presently the believe, both sets are
equal, remains a conjecture. \footnote{Thanks to the referee, who
pointed at the gap.} $\Box$

{\bf Lemma 4}

Let $H_{\omega} = 0$. Then $\omega$ is pure. $\Box$

{\it Proof}

Let us consider an arbitrary convex decomposition (\ref{conv01}).
Then, by definition of $R$ and by concavity of $S$
$$
R(\omega) \leq \sum p_j S(\tilde \omega_j) \leq S(\tilde \omega)
$$
The assumption of the lemma implies equality. But $S$ is strictly
concave. Hence $\tilde \omega$ must be equal to $\tilde \omega_j$
for all $j$. Because every state of the face of $\omega$ can
occur in a convex decomposition of $\omega$, the whole face
is reduced to a single state on ${\cal C}$ by (\ref{th02}).
For a maximal commuting subalgebra such a face cannot contain
more than one state.

Remark that the inverse statement is evident: If $\varrho$
is pure then $H(\varrho) = 0$.
$\Box$

In the following $\omega \to \bar \omega$ denotes a
complex conjugation such that $P_j (\bar \omega - \omega) P_j = 0$
for all $j$. In a suitable base for the density operators
the complex conjugation changes the off-diagonal entries to its
complex conjugates but does not change the diagonal.
If $\omega = \bar \omega$, the state is called {\it real}.

{\bf Lemma 5}

If $\varrho$ and $\bar \varrho$ both appear in an optimal
extremal decomposition then $\varrho = \bar \varrho$.

Let $U \in {\cal C}$ be a unitary. If $\varrho$ and its
transform $\varrho^U$ both appear in a proper optimal
decomposition, then they are equal.
$\Box$

{\bf Corollary}

The set of real states is an $R$-set. Every pure state occurring
in an optimal decomposition of a real state is real.
$\Box$

{\it Proof}: Let $\varrho$ be a pure state and
$\tau = (\varrho + \bar \varrho)/2$. Then $\varrho$, $\bar \varrho$,
and $\tau$ have the same reduction to ${\cal C}$ and the same
$\tilde S$-value. Assume the two extremal elements would appear in an
optimal decomposition. Then $R$ is affine on their convex hull.
(lemma 3). Hence $R(\tau) = \tilde S(\tau)$. By lemma 4
$\tau$ has to be pure implying $\varrho = \bar \varrho$.
The same chain of arguments is valid in the other case of the
lemma. $\Box$
$$     $$

{\bf USING SYMMETRIES}

If only $U^* {\cal C} \, U = {\cal C}$ is required, things are
not covered by lemma 5. These unitaries form the {\it normalizer}
of ${\cal C}$ in ${\cal B}$. They permute the minimal
projection operators $P_j$ of ${\cal C}$. Let $U$ a unitary from the
normalizer. Then there is a permutation  $i \to j(i)$ with
$U P_i U^* = P_{j(i)}$. This way we obtain the well known
homomorphism from the normalizer onto the permutation group
of $d = \dim {\cal H}$ elements. Let us call $U$
a {\it transposition} iff it interchanges two minimal projections
while the other ones remain unchanged.

My aim is to consider optimal decompositions of
states which are generated by symmetries. If $U$
is a unitary, $\omega^U$ is defined by $\omega^U(A) = \omega(UAU^*)$
for all $A$ in the algebra.
The computations are conveniently done by the help of
density operators. Using the trace of ${\cal B}$, the latter
is defined by

\begin{equation} \label{tr1}
\omega(A) = {\rm Tr} \, D \, A, \quad D = D_{\omega}, \quad \forall \, A
\end{equation}

If $\omega$ is transformed to $\omega^U$, the density operator
becomes $U^* D U$.

The rank of a state is by definition equal to the rank
of the smallest projection operator, say $Q$, satisfying
$\omega(Q) = \omega(\underline{1})$. $Q$ is called {\it support}
of $\omega$ and of the operator $D = D_{\omega}$.
We mention the equality of the rank of $\omega$ with the
dimension of the {\it supporting subspace} $Q {\cal H}$.
We shall need further

\begin{equation} \label{support}
 {\cal H}_{\omega} := Q \, {\cal H}, \quad {\cal B}_{\omega} :=
Q {\cal B} Q = {\cal B}({\cal H}_{\omega})  \cong {\cal M}_k
\end{equation}

where $k = {\rm rank} \, \omega$.
$Q$ is the unit element of ${\cal B}_{\omega}$. If
$A \in {\cal B}_{\omega}$ is of rank $k$, then $A$ is
invertible in that algebra, i.e.~there is a unique
$B \in {\cal B}_{\omega}$ with $AB = Q$. In particular,
every positive power $D^s$ is invertible in ${\cal B}_{\omega}$.
Below this will be used with $k = 2$, in which case things can
be controlled explicitly by the help of Pauli operators. Before
going to that issue, let us rewrite (\ref{th03}) with the projection
operators, $P_j$, of our maximal commutative subalgebra ${\cal C}$
$$
\tilde S(\omega) =  \sum s( {\rm Tr} \, P_j D)
$$
Consider now a {\it rank two state} $\omega$ with
density operator $D$ and support $Q$. Then

\begin{equation} \label{t-0}
{1 \over 2} \, Q = ( 1 - {\rm Tr} \, D^2 )^{-1}
( D - D^2 ), \quad
\end{equation}

{\bf Lemma 6}

Let $\omega$ be a state of rank two and $U$ be a transposition
such that $\omega^U = \omega$, so that its density operator
$D = D_{\omega}$ commutes with $U$.
Then the following properties are equivalent:

(a) There is no other $U$-invariant state in
$\Omega_{\omega}$ than $\omega$.

(b) $\omega$ allows for an optimal decomposition of $\omega$
of length two, and at least one element of $\Omega_{\omega}$
does not commute with $U$.

(c) $\Omega_{\omega}^{\rm ex}$ consists of two elements which
are interchanged by $U$.  $\Box$

{\it Proof.}
To be definite we choose a transposition $U$ fulfilling

\begin{equation} \label{t-1}
U P_1 = P_2 U, \quad U P_j = P_j U, \, \, \, \forall j > 2
\end{equation}

There is a  $180^o$-rotation in ${\cal B}_{\omega}$
through the action of $U$. (If not, all elements of that
algebra had to be $U$-invariant,
contradicting every of the three properties, a, b, c.)
We choose matrices, $\sigma_j$, in this algebra
satisfying the algebraic properties of the Pauli matrices,
with $\sigma_3$ defining the rotational axis of $U$. We are
allowed to require

\begin{equation} \label{t-3}
\sigma_3 := U \, Q,
\quad \sigma_j U + U \sigma_j = 0, \quad j = 1,2
\end{equation}

because $U$ commutes with $U$ by virtue of (\ref{t-0}), and,
being a transposition, $U = U^*$. Therefore, $\sigma_3$ of
(\ref{t-3}) is Hermitian and its square equals $Q$. But
$Q \neq UQ$ because $U$ induces a non-trivial rotation of the
supporting subspace. As we now can see, every $U$-invariant
operator in ${\cal B}_{\omega}$ is a linear combination of
$Q$ and $\sigma_3$. In particular,

\begin{equation} \label{t-4}
D = {1 \over 2} \bigl( Q + x_3 \sigma_3 \bigr), \quad
{1 \over 2}(1 + x_3^2) = {\rm Tr} \, D^2
\end{equation}

With any pure $\varrho$ also $\varrho^U$ is contained
in $\Omega_{\omega}$. Assuming property (a) of lemma 6 we obtain
the optimal decomposition

\begin{equation} \label{t-5}
{1 \over 2} ( \varrho + \varrho^U) = \omega
\end{equation}

Thus (a) $\to$ (b).

The density operator $D_{\varrho}$ of
any pure $\varrho$ satisfying (\ref{t-5}) must be of the form

\begin{equation} \label{t-6}
D_{\varrho} = {1 \over 2} \bigl( Q + \sum x_j \sigma_j \bigr),
\quad
x_1^2 + x_2^2 = 1 - x_3^2 = 2 ( 1 - {\rm Tr} \, D^2 )
\end{equation}

For $j > 2$ the projections $P_j$ commute with $U$. But
$P_j \sigma_k$, $k = 1,2,$ change sign if transformed with $U$.
Their traces must be zero. This implies

\begin{equation} \label{t-6a}
{\rm Tr} \, P_j D = {\rm Tr} \, P_j D_{\varrho}, \quad
j > 2
\end{equation}
\begin{equation} \label{t-6b}
{1 \over 2} \, {\rm Tr} \,(P_1 + P_2) D_{\varrho} = {\rm Tr} \, P_1 D
= {\rm Tr} \, P_2 D
\end{equation}

In order that (c) follows from (b) there should be only two
choices for $\varrho$ if optimality is required. By (\ref{t-6a})
and (\ref{t-6b}) the remaining possibility to get $\tilde S(\varrho)$
as small as possible is in making the modulus of the difference
${\rm Tr}(P_1 - P_2)D_{\varrho}$ as large as possible. To do this
is the next aim.

If the above mentioned difference is zero, then
${\rm Tr} P_j D_{\varrho} = {\rm Tr} P_j D$ for all $j$.
But then the entropy of $\tilde \varrho$ and $\tilde \omega$ would be
equal, and, consequently, $H_{\omega} = 0$.
By lemma 4 this contradicts the rank two
assumption for $\omega$. Therefore, the first of the operators
$$
D_{\varrho}(P_1 - P_2)D_{\varrho}, \quad
Q(P_1 - P_2)Q, \quad \sqrt{D} (P_1 - P_2) \sqrt{D}
$$
is not zero. (Remark $D_{\varrho} = \sqrt{D_{\varrho}}$, for $\varrho$
is pure.)
But the first operator results from multiplying the second one
from the right and the left by $D_{\varrho}$. Thus the
second operator is again not the zero of ${\cal B}_{\omega}$.
And, lastly, the root of $D$ is invertible in ${\cal B}_{\omega}$.
This means the third operator is not the zero. All the
operators of our list are Hermitian.
All change sign if transformed with $U$. Hence they are
real linear combinations of $\sigma_1$ and $\sigma_2$. We use
the remaining freedom in the choice of the Pauli operators by
requiring

\begin{equation} \label{t-7}
Q \, (P_1 - P_2) \, Q = y \, \sigma_1, \quad y > 0
\end{equation}

Multiplication with $D_{\varrho}$ and taking the trace gives

\begin{equation} \label{t-8}
{\rm Tr} \,(P_1 - P_2) D_{\varrho} =  x_1 y
\end{equation}
according to (\ref{t-6}). The left hand side is maximal
(minimal) iff $|x_1|$ is as large as possible.
This takes place if and only if $x_2=0$, see (\ref{t-6}).

That proves (b) $\to$ (c).

Indeed, with $x_2 = 0$, just one choice for $x_1 > 0$ is allowed,

\begin{equation} \label{t-11}
\Omega_{\omega}^{\rm ex} = \{ \, \varrho, \varrho^U \, \},
\quad D_{\varrho} = {1 \over 2} \bigl( Q + x_1 \sigma_1 +
x_3 \sigma_3 \bigr) = D + {1 \over 2} \sqrt{1 - x_3^2} \sigma_1
\end{equation}

Evidently, (a) follows from (c), and the proof is done. $\Box$

Let us register the following observation:

{\bf Corollary}

Let $\omega$ satisfy the conditions of Lemma 6. The choice
(\ref{t-3}), (\ref{t-7}) of the Pauli operators $\sigma_j$
depends {\it only} on the support $Q$ of $\omega$. $\Box$

To see it, remind that $P_j$ are rank one projections and that
the traces of $P_1Q$, $P_2Q$ are equal. We get $y$ by
squaring (\ref{t-7}) and taking the trace. Reinserting
in (\ref{t-7}) yields

\begin{equation} \label{t-12}
\bigl( \sqrt{ {\rm Tr}(P_1 Q) \,{\rm Tr}(P_2 Q) -
{\rm Tr}(P_1 Q P_2 Q)} \, \bigr) \, \sigma_1 = Q (P_1 - P_2) Q
\end{equation}

Similarly we see from (\ref{t-6}) and by sandwiching (\ref{t-7})
with $\sqrt{D}$

\begin{equation} \label{t-13}
\sqrt{D} (P_1 - P_2) \sqrt{D} =  {1 \over 2} x_1 y \sigma_1
\end{equation}

Squaring and taking the trace comes down to

\begin{equation} \label{t-14}
{1 \over 4} \, x_1^2 y^2 =
{\rm Tr}(P_1 D) \,{\rm Tr}(P_2 D) - {\rm Tr}(P_1 D P_2 D)
\end{equation}

To be able to calculate $\tilde S(\varrho) = R(\omega)$
we need the traces of $D_{\varrho}P_j$, $j=1,2$. This can
be done by combing (\ref{t-6b}) and (\ref{t-8}):

$$
{\rm Tr} \,P_1 D_{\varrho} = {\rm Tr} (P_1 D) + (x_1y)/2, \quad
{\rm Tr} \,P_2 D_{\varrho} = {\rm Tr} (P_1 D) - (x_1y)/2
$$

Now, because of (\ref{t-14}), explicit expressions for $R$ and
$H_{\omega}$ are

\begin{equation} \label{t-15}
R(\omega) = s({\rm Tr}(P_1 D) + {1 \over 2} x_1y) +
s({\rm Tr}(P_1 D) - {1 \over 2} x_1y)
+ \sum_{j>2} s({\rm Tr} P_j D)
\end{equation}

\begin{equation} \label{t-16}
H_{\omega} = 2 s({\rm Tr}(P_1 D) -
s({\rm Tr}(P_1 D) + {1 \over 2} x_1y) -
s({\rm Tr}(P_1 D) - {1 \over 2} x_1y)
\end{equation}

{\it provided the assumption of lemma 6 are satisfied.}
The trace of $P_jD$ is the expectation value of $D$
with the vector $|j\rangle$. Similarly, (\ref{t-14})
may be written

$$
{1 \over 4} x_1^2 y^2 =
\langle 1| D |1 \rangle \langle 2| D |2 \rangle -
\langle 1| D |2 \rangle \langle 2| D |1 \rangle
$$

Looking at all this, the difficulties in extending lemma 6 to
symmetric density operators of higher rank are as follows:
(\ref{t-0}) becomes an equation of degree rank$(\omega)$, and
the number of parameters goes quadratically with the rank.

There is a remarkable outcome of lemma 6. With an arbitrary pure
state $\varrho$ and a given transposition $U$ there is a
twofold alternative. At first, either $\varrho = \varrho^U$ or
the arithmetic mean (\ref{t-5}) of $\varrho$ and
$\varrho^U$ is of rank two. In the latter case $\Omega_{\omega}$
is $U$-invariant. Hence, either the conditions
of lemma 6 are satisfied or they are not. In the latter case, $\varrho$
is not optimal. However, because of the symmetry, there is
necessarily at least one optimal
pure $\varrho_1$ in $\Omega_{\omega}$ such that lemma 6 applies
to the arithmetic mean of $\varrho_1$ and $\varrho_1^U$.
Hence $\Omega_{\omega}^{\rm ex}$ consists of
one or more pairs of pure states, which are pairwise
permuted by $U$, and, possibly, of some $U$-invariant pure states.

{\bf Corollary}

If $\omega$ is $U$-invariant but $\Omega_{\omega}$ is not elementwise
$U$-invariant then $\Omega_{\omega}$ contains at least one state
to which lemma 6 applies.
$$    $$

{\bf SYMMETRIC REAL DENSITY OPERATORS}

Let us compare the treatment above with that of some highly symmetric
density operators of maximal rank according to \cite{BNU96}.
Assuming $\omega$ {\it real},
every optimal decomposition of $\omega$ is real (lemma 5).
Even more essential, $D = D_{\omega}$, the density operator
of $\omega$, is supposed to commute with all permutation matrices,
$U$, fulfilling $U{\cal C}U^* = {\cal C}$. In the following
the latter assumption is {\it always} requiered.

To every permutation, $\pi$, there is a unique permutation matrix,
$U_{\pi}$, in the normalizer of ${\cal C}$. These matrices
are real unitaries with entries 0 or 1, and in every row and
every column there is just one 1.
If a real density operator is commuting with all real
permutation matrices, only one free parameter remains:
It is the common value, $z$, of the
off-diagonal elements. The diagonal elements equal $1/d$,
$d$ the dimension of our Hilbert space.
The common off-diagonal value is bounded from above by $1/d$
and from below by $-1/d(d-1)$.

Now let a pure state $\varrho$ with density operator
$D_{\varrho}$ appear in an optimal decomposition
of a real permutation invariant state. Then every transform
$\varrho^U$ of $\varrho$ by a permutation matrix
is contained in
$\Omega_{\omega}^{\rm ex}$. Therefore, lemma 3 shows
optimality of the decomposition (which is not necessaily short)

\begin{equation} \label{sn-1}
D = {1 \over d !} \sum_{\pi} U_{\pi} D_{\varrho} \, U_{\pi}^*
\end{equation}

One of the relations following from (\ref{sn-1}) reads

\begin{equation} \label{rels1}
{\rm Tr} D_{\varrho} D = {\rm Tr} D^2 =
{1 \over d} + d (d-1) \, z^2
\end{equation}

We may write

\begin{equation} \label{rels1a}
D_{\varrho} = | \varphi \rangle \langle \varphi |
\end{equation}

with a real unit vector $\varphi$. Denoting by $\phi_1, \dots, \phi_d$
the components of $\varphi$ in a base that diagonalizes the minimal
projections $P_j$ of ${\cal C}$, the relation (\ref{rels1}) implies

\begin{equation} \label{rels2}
\sum \phi_k = a, \quad  a = \sqrt{1 + z d (d-1)} \geq 0
\end{equation}

where the sign of the real $a$ is fixed by
$a \geq 0$. This seemingly harmless convention has an important
effect. Being real, $\varphi$ is defined by (\ref{rels1a})
up to a sign. If $a \neq 0$, this sign has been fixed by (\ref{rels2}).
(\ref{rels2}) is an affine hyperplane, intersecting the $(d-1)$-sphere
spanned by the real unit vectors $\varphi$.
As long $a > 0$ the map $D_{\varrho} \to \varphi$ is a section
from the real pure states into the Hilbert space. For
$a = 0$ we get a double covering because with $\varphi$ also
$-\varphi$ belongs to the sphere. That is, in the limit $a \to 0$
the simple covering bifurcates to a double covering.

The point for all this comes from lemma 3, showing that
(\ref{sn-1}) implies $R(\omega) = S(\tilde \varrho)$ because $R$
is affine on the convex set generated by an optimal set of
pure states.

Thus we have to minimize
$S(\tilde \varrho)$ on the intersection of the $(d-1)$-sphere
of real unit vectors with the hyperplane (\ref{rels2}),
i.e.~on a $(d-2)$-sphere ${\cal S}_a^{d-2}$. Its radius $r$
in Hilbert space turns out to be

\begin{equation} \label{radius}
r = r( {\cal S}_a^{d-2} ) = \sqrt{1 - {a^2 \over d}}
= \sqrt{{(d-1)(1 - zd) \over d}}
\end{equation}

From $z= 1/d$, where it degenerates to a point, the radius grows
up to one with growing $z$. At the same time $a$ goes from
$\sqrt{d}$ to zero.

Let $\varphi \in {\cal S}_a^{d-2}$ and denote by $\varphi^{\perp}$
its antipode on that sphere. Then their Hilbert distance
is twice the radius (\ref{radius}), which amounts to

\begin{equation} \label{trp}
\langle \varphi, \varphi^{\perp} \rangle = 2 {a^2 \over d} -1
= 1 - 2 r^2
\end{equation}

so that the transition amplitude remains positive as long as the
radius does not exceed $r_0 := \sqrt{0.5}$.
Thus, for $0 \leq r \leq r_0$,
the Bures distance of the states $|\varphi \rangle\langle \varphi|$
and $|\varphi^{\perp} \rangle\langle \varphi^{\perp}|$
is equal to the Hilbert distance of $\varphi$ to its antipode
$\varphi^{\perp}$. But for $r_0 < r < 1$ the transition amplitude
becomes negative and the mentioned Bures distance gets the value
$2 \sqrt{1 - r^2}$. This can be rephrased as following:
Within $0 \leq r < 1$ the sphere ${\cal S}_a^{d-2}$ is one
to one mapped into the state space. This mapping is locally
isometric. The local isometry is a global one for
$0 \leq r \leq r_0$. But it becomes globally deformed if $r$
is larger than $r_0$ in order to ``prepare'' the bifurcation
at $r=1$.
Because of the described scenario something should happen
with the optimization and its outcome $R$. What it is, is
definitely known \cite{BNU96} in case $d=3$, and will be
described below.

For the next considerations
I assume $d = 3$. With $d-2 = 1$ the optimalization takes
place on an 1-sphere. There are three permutation matrices which
are transpositions. They are denoted by $U_{12}$, $U_{23}$, and
$U_{31}$. In particular, the real unitary $U_{12}$
interchanges the components $\phi_1$ and $\phi_2$ of $\varphi$,
while $\phi_3$ remains unchanged, and so forth.
The product of any two of the three
transpositions is a cyclic permutation of the components of $\varphi$.

Now I return to an important result of \cite{BNU96} which
clarifies the structure of $\Omega_{\omega}^{\rm ex}$ in its
dependence on $z$.

There is a special $z$-value, the {\it bifurcation value} $z_*$,
which is $-0.14$ approximately.
For values $-1/6 < z < z_*$ the convex set $\Omega_{\omega}^{\rm ex}$
is an {\it hexagon}. But for
$z_* \leq z < 1/3$ it is a {\it triangle}, i.e.~a {\it simplex}.

In the {\it triangle} case there is, up to reordering of its
extremal states, exactly one short optimal
decomposition of $\omega$. It is of length three and explicitly known
\cite{BNU96}.  The following representation of their density matrices
$D_j$ is in \cite{Ben96}. It is

\begin{equation} \label{sn-t}
\omega = {1 \over 3} \, ( \varrho_1 + \varrho_2 + \varrho_3 ),
\quad D_j := 3 \, \sqrt{\omega} P_j \sqrt{\omega}
\end{equation}

This representation is equivalent to (\ref{sn-1}): The $6 = 3!$
terms in (\ref{sn-1}) become pairwise equal.
Every transposition permutes two of the three pure states
in $\Phi_{\omega}$, allowing for an application of lemma 6,
but let the third one unaffected. On the other hand, a
cyclic permutation matrix induces a
cyclic reordering of $\Omega_{\omega}^{\rm ex}$, i.~e.~of
three pure states of (\ref{sn-t}).

More involved is the {\it hexagon} case.
After the bifurcation value every of the three optimal pure states
of the simplicial decomposition splits into two other ones.
That is to say, from every one of the three pure states $\varrho_j$,
$j = 1,2,3$ of the triangular optimal decomposition
originates two new ones, $\varrho_{ja}$ and $\varrho_{jb}$.
A transposition, say $U_{12}$,
previously interchanging $1 \leftrightarrow 2$ but letting
$3$ fixed, now does a more complicated job:
$1a \leftrightarrow 2b$,
$1b \leftrightarrow 2a$, and $3a \leftrightarrow 3b$.
The pair $\varrho_{3a}$, $\varrho_3b$, together with
$U_{12}$ allows for the application of lemma 6.

The states labelled by $a$ are interchanged by a cyclic
permutations, and the same is with the $b$-states $\varrho_{bj}$.
From that one obtains two essentially different
simplicial decompositions,

\begin{equation} \label{sn-2h}
\omega = {1 \over 3} ( \varrho_{1a} + \varrho_{2a} + \varrho_{3a} )
= {1 \over 3} ( \varrho_{1b} + \varrho_{2b} + \varrho_{3b} )
\end{equation}

However, no pair with the same index ``$a$'' satisfies the assumptions
of lemma 6. The same is with respect to the index ``$b$'':
Only states on the line segment containing two neighbored extremal
points can have a {\it unique} extremal decomposition.

We already know:
Something appears if $z$ goes down to $-1/6$. The
Study-Fubini distance of the pairs of pure states labelled by
$(1a, 2b)$, $(2a, 3b)$, or $(3a, 1b)$ respectively, is
diminishing. The distance finally becomes zero for
$z = -1/6$, resulting in $\varrho_{1a} = \varrho_{2b}$,
and so on. The hexagon bifurcates and becomes again a triangle
in state space.

In the Hilbert space one gets an equilateral hexagon at
$z = -1/6$.
As explained above, it becomes our triangle in state space
by identifying the vectors $\varphi$ and $-\varphi$ (Hopf bifurcation).
From this point of view it really looks as if we
had to compensate that Hopf bifurcation by the bifurcation
of the optimal decomposition rule at $z_*$.
If this impression is correct, the appearance of $z_*$
is necessary by general geometric reasons.  Only its
value should come from the particular properties of
the function $s(x)$.

Let us compare this reasoning again with lemma 6.
Let $\varrho$ be a real pure state,
$U$ a real transposition that does not commute with $\varrho$,
and denote by $\omega'$ their arithmetical mean,
$\omega' = (\varrho + \varrho^U)/2$. If the latter is not an
optimal decomposition, then $\Omega_{\omega'}$ is spanned by
more than two extremal states. The assumption, that we then fall
into the triangle or hexagon case, is compatible with the
symmetry and geometrically tempting. This conjecture reads:

Let $\varrho$ a real pure state, $U$ be a real transposition,
and $\varrho \neq \varrho^U$. If the assumptions of lemma 6
do not apply to the state $(\varrho + \varrho^U)/2$, then
$\varrho$ belongs to an optimal decomposition of a real and
maximally symmetric state. $\Box$

The conjecture is supported by numerical studies and the results
of \cite{BNU96}. To get a complete proof, one has to exclude
further bifurcations. I do not know how to achieve this. $\Box$

{\it Remark:} For $d = 2$ every pair $\varrho$, $\varrho^U$
of pure states, $U$ a transposition of the minimal projections
of ${\cal C}$ is optimal. Indeed, this remains true if $-x \ln x$
is replaced by an arbitrary concave $s(x)$ with $s(0) = s(1) = 0$. $\Box$

For $d > 3$ a similar analysis is preliminary only.
To obtain a pure state $|\varphi \rangle \langle \varphi|$ belonging
to an optimal decomposition (\ref{sn-1}) it suffices to restrict
oneself to the following assumption: The components of $\varphi$
do not attain more than three different values. This can be shown
by straightforward variational analysis.

Moreover, if the components of $\phi$ attain only two
different values, $\varrho$ is either a local minimum, a maximum,
or a turning point of $R$. Nearby $z = 1/d$ the vector $\varphi$
with components

\begin{equation} \label{sn-3}
\sum \phi_j \geq 0, \quad \phi_1 > \phi_2 = \phi_3 = \dots = \phi_d
\end{equation}

gives at least a local minimum of
$S(\tilde \omega)$ which is presumably a global one.
The $U_{\pi}$-transforms of $D{\varrho} =
|\varphi \rangle \langle \varphi|$, where $\varphi$ satisfies
(\ref{sn-3}), generate a simplex spanned by $d$ extremal states.
For $z$-values satisfying  (\ref{rels2}) and (\ref{sn-3}),
and such that (\ref{sn-1}) becomes optimal (though not short),
the simplex decomposition will be

\begin{equation} \label{sn-nt}
\omega = {1 \over d} \sum \varrho_j, \quad
D_j := d \, \sqrt{\omega} \, P_j \, \sqrt{\omega}
\end{equation}

This is supposed to be the counterpart, for $d > 3$, of the
$d = 3$ triangle case.
Of course, much more has to be known to clarify the $d>3$ situation
even in the real and maximally symmetric case.

What is to learn about the role of symmetries from all that?
Given a state $\omega$, one is tempted to look at
the subgroup

\begin{equation} \label{ngroup1}
\Gamma(\omega) := \{ \, U \, | \, \, U^* {\cal C} U = {\cal C},
\, \, \, U^* \Omega_{\omega} U = \Omega_{\omega} \,   \}
\end{equation}

of the normalizer of ${\cal C}$. As seen in the previous examples,
a certain classification can be reached
by examining to the detail the action of $\Gamma(\omega)$ on
the pure states contained in $\Omega_{\omega}^{ex}$.
Is it always true, as in the examples considered above, that
{\it there is exactly one} $\omega' \in \Omega_{\omega}$
{\it which is invariant with respect to} $\Gamma(\omega)$ ?
And, if yes, is this group large enough to get an optimal
decomposition of $\omega'$, starting with any of its optimal
pure states?
$$   $$

{\bf Appendix: \quad Roofs}

A function enjoys some very nice properties if
defined according the rule of (\ref{conv02}). Some of them
have been used by Benatti, Narnhofer, and Uhlmann \cite{BNU96},
by Uhlmann \cite{Uh97a}, by Hill and Wootters \cite{HW97}, and
others to examine either the {\it entropy of a channel or of
a subalgebra with respect to a state}, \cite{CNT87}, or the
{\it entanglement of formation}, \cite{BDSW96}.
They can also provide computational help to Holevo's channel
capacity \cite{HolCap}. In addition there are
connections to the optimalization problem of
accessible entropy shown by Benatti \cite{Ben96}. They explain
certain  similarities to results of Davies \cite{Da78},
Levitin \cite{Lev94}, Fuchs and Peres \cite{FP96}.

In the following I treat these
general properties within an abstract setting.
Its first requirement is as follows:

ASSUMPTION 1: $\Omega$ {\it is a compact convex
set in a finite dimensional real space} ${\cal L}$.

{\it Remark:} In most
physical applications $\Omega$ is the state space or the space of
density operators of an algebra ${\cal B}(\cal H)$, $^*$-isomorph
to a full matrix algebra. ${\cal H}$ denotes an Hilbert-space
of finite dimension $d$. Fixing $\Omega$ to be the convex set
of all density operators, ${\cal L}$ is the real linear space of
Hermitian operators, Herm$({\cal H})$, of ${\cal H}$. Then
${\cal L}$ is of dimension $d^2$. The dimension of $\Omega$
is $d^2 - 1$. Only for reference within the appendix I call this
{\it the standard setting}. It is convenient to require

\begin{equation} \label{A1-1}
n := \dim {\cal L} = 1 + \dim \Omega, \quad 0 \notin \Omega
\end{equation}

This provides the following: Because the zero of ${\cal L}$
is not contained in $\Omega$, the linear span of $\Omega$
coincides with ${\cal L}$. Choose $\tau \in \Omega$ arbitrarily.
To every $\nu \in {\cal L}$ there is one and only one real
number $\lambda$ such that $\nu - \lambda \, \tau \in \Omega$.
For the remainder $\tau$ is chosen once for all as a
{\it reference state}. It is often convenient, though not
necessary, to assume invariance of $\tau$ against all
affine automorphisms of $\Omega$. ($\tau$ is then called
maximally symmetric.)

ASSUMPTION 2: {\it The set of extremal elements, $\Omega^{\rm ex}$,
of $\Omega$ is compact. A continuous function, $\varrho \to f(\varrho)$,
is given on $\Omega^{\rm ex}$.} $\Box$

The next aim is to extend the function given on the extremal boundary
of $\Omega$ to the whole convex set $\Omega$. Of course, there are
many ways to do so. But there exists two distinguished among
them, respecting maximally the convex structure of $\Omega$.
For reasons which will became evident soon, I call them the
{\it the convex} and {\it the concave roof based on} $f$.
The {\it convex roof}, $f^{\inf}$, is defined by

\begin{equation} \label{A1-2}
f^{\inf}(\omega) := \inf \sum p_j f( \varrho_j ),
\quad \varrho_k \in
\Omega^{\rm ex}, \quad \sum p_j \varrho_j = \omega
\end{equation}

where the infimum runs through all extremal convex decompositions
of $\omega$. Completely similar, the {\it concave roof},
$f^{\sup}$, is defined by

\begin{equation} \label{A1-3}
f^{\sup}(\omega) := \sup \sum p_j f( \varrho_j ),
\quad \varrho_k \in
\Omega^{\rm ex}, \quad \sum p_j \varrho_j = \omega
\end{equation}

Because $-f^{\sup} = (-f)^{\inf}$ every property of
convex roofs can be translated in one for concave roofs, and vice
versa. Evidently $f^{\sup} \geq f^{\inf}$.

The task is now to show how the graphs of $f^{\sup}$ and
$f^{\inf}$ unite to the boundary of a compact convex
set $\Xi$ of dimension $n$. It will be done by a construction
depending on the reference state $\tau$. The set

\begin{equation} \label{A1-4}
\Xi^{\rm ex} := \{ \, f(\varrho) \, \tau + \varrho \, |
\quad \varrho \in \Omega^{\rm ex} \,   \}
\end{equation}

does not contain convex linear combinations of their elements
with the exception of the trivial ones. Otherwise $\Omega^{\rm ex}$
could not be a set of extremal points of a convex set.
Continuity of $f$ and compactness of $\Omega^{\rm ex}$ imply
compactness of $\Xi^{\rm ex}$. Hence (Carath\'eodory)

{\bf Lemma A-1}

The convex hull $\Xi$ of $\Xi^{\rm ex}$ is compact.
$\Xi^{\rm ex}$ is the set of extremal points of $\Xi$. $\Box$

$\nu \in \Xi$ iff $\nu$ allows for an extremal decompositon
$$
\nu = [ \sum p_j f(\varrho_j) ] \tau + \sum p_j \varrho_j
$$
resulting in

\begin{equation} \label{A1-5}
\lambda \, \tau + \omega \in \Xi \quad \longleftrightarrow
\quad  f^{\inf} \leq \lambda \leq  f^{\sup}
\end{equation}

The compactness of $\Xi$ ensures the compactness of the
$\lambda$-interval defined by (\ref{A1-5}). It follows the
existence of {\it optimal decompositions} with which the
``inf'' in (\ref{A1-2}) or the ``sup'' in (\ref{A1-3}) are
attained respectively.
Moreover, $\lambda \tau + \omega$
belongs to the boundary of $\Xi$ iff $\lambda$ equals either
$f^{\inf}(\omega)$ or $f^{\sup}(\omega)$. The dimension
of its face cannot exceed $n-1$. Carath\'eodory's theorem
guaranties optimal decompositions of length $n$.

{\bf Lemma A-2}

$f^{\inf}$ as well as $f^{\sup}$ allows for optimal
decompositions of length not exceeding $n$. $\Box$

There is another construction, \cite{Roc70}, valid on every convex
set. Let $\omega \to G(\omega)$ denote an arbitrary function
on $\Omega$. Its {\it convex hull} is defined by

\begin{equation} \label{A1-6}
G_{\inf}(\omega) := \inf \sum p_j G( \omega_j ),
\quad \omega_k \in
\Omega, \quad \sum p_j \omega_j = \omega
\end{equation}

where the inf runs through {\it all} representations of $\omega$ by
convex combinations, i.~e.~not necessarily extremal ones. It is
known, see \cite{Roc70}, and easily verified, that {\it every convex
hull is a convex function}.

In the same spirit one my define the {\it concave hull} of a function
just by replacing the ``inf'' in (\ref{A1-6}) by ``sup''. With this
definition the concave hull of a function is concave.

Now assume in (\ref{A1-6}) a concave function $G$. Then we may restrict
ourselves to extremal decompositions to get the desired infimum:
The convex hull of a concave function depends on it values at
the extremal points only. Now it is straightforward to see

{\bf Lemma A-3}

$f^{\inf}$ is the convex hull of $f^{\sup}$.
$f^{\inf}$ is convex.
Let $F$ be a convex function which is not larger than $f$ on
$\Omega^{\rm ex}$. Then $F \leq f^{\inf}$ on $\Omega$.

$f^{\sup}$ is the concave hull of $f^{\inf}$.
$f^{\sup}$ is concave.
Let $F$ be a concave function which is not smaller than $f$ on
$\Omega^{\rm ex}$. Then $F \geq f^{\sup}$ on $\Omega$.
$\Box$

{\it Remark A1.2:} Let us consider the standard setting where
$\Omega$ is a state space and $S$ the von Neumann entropy.
If $\omega \to \omega \circ \alpha$ denotes an affine mapping
of the state space into itself then
$$
\omega \mapsto \tilde S(\omega) := S( \omega \circ \alpha )
$$
is a concave function on $\Omega$. Its convex hull,
$\tilde S_{\inf}$, is
denoted by $R$ in \cite{BNU96} and \cite{Uh97a} and by $E$
for ``entanglement'' in \cite{BDSW96} and \cite{HW97}. Let
$f$ denote the restriction of $\tilde S$ onto $\Omega^{\rm ex}$.
We have
$$
\tilde S \geq f^{\sup} \geq f^{\inf} = \tilde S_{\inf}, \quad
H_{\omega} = \tilde S - \tilde S_{\inf} \geq f^{\sup} - f^{\inf}
$$
Because all the functions are non-negative, they are defined
for general state spaces (say in the C$^*$-category) if the
von Neumann entropy remains finite on $\Omega \circ \alpha$.
$\Box$

Now a further notation is introduced. Let $F$ be a function
on $\Omega$. A set of extremal points of $\Omega$ is called
$F$-{\it optimal} if and only if $F$ is affine on its convex hull.

I call $F$ a {\it roof} if every element $\omega$
is contained in the convex hull of an $F$-optimal set.

This is consistent with the notations above:
The concave and the convex roof of a function $f$,
which is defined on the extremal points, are roofs.
It is the content of
theorem 1 in \cite{BNU96}. As already indicated in the main text
(lemmata 1-3), one can do a little bit more. What there is called
$\Omega_{\omega}$ will now be denoted by $\Omega^{\inf}_{\omega}$
to distinguish it from $\Omega^{\sup}_{\omega}$.

$\Omega^{\inf}_{\omega}$ is the smallest convex set
containing all extremal points of all extremal decompositions
of $\omega$ which are optimal for $f^{\inf}$.

$\Omega^{\sup}_{\omega}$ is the smallest convex set
containing all extremal points of all extremal decompositions
of $\omega$ which are optimal for $f^{\sup}$.

{\bf Lemma A-4}

$f^{\inf}$ and $f^{\sup}$ are roofs. They are affine
on $\Omega^{\inf}_{\omega}$ and $\Omega^{\sup}_{\omega}$
respectively.

{\bf Corollary}

The convex hull of a concave function and the concave hull of
a convex function are roofs.

Two convex (or two concave) roofs are equal iff they coincide
on the extremal points.      $\Box$

The proofs are mere reformulations of those in the main text. They
can be done also more explicitly as in \cite{BNU96}.

Remark: We may now rephrase the definition of $H_{\omega}$
as following: $H_{\omega}$ vanishes on $\Omega^{\rm ex}$
and it is the sum of $\tilde S$ and of a concave roof. $\Box$

What happens if equality holds, $f^{\sup}(\omega) = f^{\inf}(\omega)$,
for a certain $\omega$. From the very definition
$$
f^{\inf}(\omega) \leq \sum p_j \varrho_j \leq f^{\sup}(\omega)
$$
so that {\it every} extremal decomposition is optimal for both
roofs. That implies $\Omega^{\inf}_{\omega}$ is the face of
$\omega$ in $\Omega$. Now the roof property implies:

{\bf Lemma A-5}

Let $f^{\sup}(\omega) = f^{\inf}(\omega)$. Then

\begin{equation} \label{A1-7}
\Omega^{\inf}_{\omega} = \Omega^{\sup}_{\omega} = \hbox{face of }
\omega \hbox{ in } \Omega
\end{equation}

and $f^{\inf}$ is equal to $f^{\sup}$ on the face of $\omega$.


{\bf Acknowledgement.} I like to thank Heide Narnhofer, Peter Alberti,
Fabio Benatti, Bernd Crell, and Christopher Fuchs for valuable
discussions and correspondence.


\end{document}